\documentclass[aip,apl,reprint,floatfix]{revtex4-1}
\usepackage{lineno,hyperref}
\usepackage{graphicx,float} 
\usepackage{bm}      
\usepackage{gensymb}
\usepackage{geometry}
\usepackage{amsmath}
\usepackage{color}
\usepackage{esvect}
\usepackage{natbib}
\begin{document}
\title{Pushing the limits of atomistic simulations towards ultra-high temperature: a machine-learning 
force field for ZrB$_2$}
\author{Yanhui Zhang, Alessandro Lunghi, Stefano Sanvito}
\address{School of Physics and CRANN, Trinity College, Dublin 2, Dublin, Ireland}

\begin{abstract}
Determining thermal and physical quantities across a broad temperature domain, especially up to the ultra-high 
temperature region, is a formidable theoretical and experimental challenge. At the same time it is essential for 
understanding the performance of ultra-high temperature ceramic (UHTC) materials. Here we present the 
development of a machine-learning force field for ZrB$_2$, one of the primary members of the UHTC family 
with a complex bonding structure. The force field exhibits chemistry accuracy for both energies and forces 
and can reproduce structural, elastic and phonon properties, including thermal expansion and thermal transport. 
A thorough comparison with available empirical potentials shows that our force field outperforms the competitors. 
Most importantly, its effectiveness is extended from room temperature to the ultra-high temperature region (up to $\sim$2,500~K), 
where measurements are very difficult, costly and some time impossible. Our work demonstrates that machine-learning force 
fields can be used for simulations of materials in a harsh environment, where no experimental tools are
available, but crucial for a number of engineering applications, such as in aerospace, aviation and nuclear.
\end{abstract}

\maketitle

\section{Introduction}
Transition metal borides, carbides and nitrides, such as ZrB$_2$, HfB$_2$, ZrC and TiN, are formed by combining a light
element from group III, IV or V, with a refractory transition metal. They are generally characterized by an unusual
combination of physical and chemical properties, such as an extremely high melting temperature ($T>$~3000~$\degree$C),
great hardness, and both good chemical stability and mechanical strength at high temperatures~\cite{uhtc}. These
compounds are usually referred to as ultra-high-temperature ceramics (UHTCs). They are considered to be the most relevant
materials set for high-temperature applications in the aerospace (e.g. thermal protection systems for hypersonic or 
atmospheric reusable re-entry vehicles, propulsion components, combustion chambers, engine intakes or rocket
nozzles), energy (e.g. nuclear fission and fusion, energy harvesting, concentrated solar power) and high-technology (e.g.
high temperature electrodes, high speed machining tools, molten metal containment) sectors.~\cite{Caccia2018,Padture2016,Zeng2017,Fahrenholtz2017}

UHTCs containing either ZrB$_2$ or HfB$_2$ are possibly the most widely studied members of this class due to their additional 
resistance to oxidation observed up to 2000~$\degree$C. In particular, they hold expectations as a replacement of silicon-based 
compounds, such as SiC, Si$_3$N$_4$ and MoSi$_2$, whose operation range is limited to approximately 1700~$\degree$C. 
Notably, the hot section of aerospace and nuclear facilities can easily approach working temperatures of 2000~$\degree$C in 
a very short time (0.1 second to a couple of minutes), meaning that the typical thermal shocks are considerable. Clearly, the 
knowledge and control of the thermal and physical properties with temperature is crucial for optimizing and understanding
the performances of these engineering materials over their entire operation range.

Ideally, it would be important to study these materials' properties in the conditions found in their service
environment~\cite{CSANADI2019}. This is particularly relevant for temperature effects, since the knowledge of
$T$-dependent properties allows one to accurately assess the impact of thermal shock, creep and fatigue, and to understand
the sensitivity of the possible engineering components to thermal-stress-induced failures. Unfortunately, the experimental
characterization of high-temperature properties is very challenging and costly, if not impossible~\cite{Ghidini2018}, so
that only an incomplete picture is captured in most cases. This means that the engineering design is somehow blind to the
most extreme part of the operation conditions. All these difficulties make theoretical simulations a really valuable
toolset to study the material's responses to such extreme conditions. Unfortunately, a universally accurate
and efficient simulation method is still not available.

{\it Ab-initio} molecular dynamics can manage high temperatures in a natural fashion, but its high computational cost
makes it practical for the simulation of only a few hundred atoms for up to hundreds of picoseconds. This is usually not
sufficient, since the required simulation scale needed to solve many problems of interest goes far beyond thousands of
atoms and several thousands of nanoseconds or even microseconds. In contrast, empirical force fields can extend to the
desired time and length scales. These are computationally efficient due to the analytical functions employed to express
the total energy and the forces. However, they usually struggle to handle several features specific to the high-temperature 
domain, namely i) the large variation of physical properties across a broad temperature range; ii) the reliability of the 
extrapolated data from empirical parameters; iii) the strong anharmonic effects taking place at 
ultra-high temperature, especially close to the melting point ($T_m\sim$~3000$\degree$C for UHTCs). 

A promising solution to this conundrum is to use machine-learning force fields (MLFFs). They can reach
quantum-chemistry accuracy at a computational cost comparable with that of conventional effective potentials. 
The formulation and implementation of MLFFs are rapidly becoming an active research area, and the latest generation 
of MLFF has also a great potential for the investigation of molecules and solids~\cite{Lunghi2019}, of amorphous
phases~\cite{Sosso2018} and of multi-component materials~\cite{Artrith2017,Rostami2018}. 

In this work we take the spectral neighbor analysis potential (SNAP) formulation~\cite{Thompson2015316} to derive 
a robust MLFF for ZrB$_2$, a prototypical UHTC material. ZrB$_2$ has the AlB$_2$ structure (P6$_3$/mmc) 
with a nano-laminate stacking of B and Zr layers. The B layer has a 2D graphene-like structure with strong 
covalent bonds, while the Zr-layer is dominated by metallic bonding. In addition, the Zr-B interaction has a 
mixed covalent and ionic nature~\cite{Vajeeston2001}. Previous attempts to construct atomic potentials for ZrB$_2$
include Tersoff potentials~\cite{Daw2011} and reactive force-fields (ReaxFF)~\cite{Gouissem2013}. These,
however, struggle with the $T$-dependent thermo-physical properties. For instance, they run into problems when 
calculating the thermal conductivity at 1000 K~\cite{Lawson2011} and they are not accurate in describing 
room-temperature elasticity. These difficulties are attributed to the inability to simultaneously describe: 
i) the rather chemically different constituent elements, ii) the complex bonding nature comprising a mixture of metallic, 
covalent and ionic interactions, iii) the intricate response to a wide range of heat and mechanical loads. 

By tuning the SNAP parameters from a rich variety of reference configurations, our derived MLFF demonstrates 
a surprisingly accurate ability to face the aforementioned challenges. A brief description of the SNAP formalism, 
the training datasets and the procedure to determine the model parameters is presented in the following section.  
Thereafter we apply the developed MLFF to predict the fundamental physical properties of ZrB$_2$. The results are
compared with available experimental and simulation findings. Our SNAP potential exhibits robustness in reproducing
the $T$-dependent physical properties and a good transferability across a wide range of temperatures and deformations.
This work provides a framework to study the high-temperature behavior of materials and it paves the way for accurate
atomistic simulations of UHTCs. 

\section{Construction of the force field}
\subsection{The SNAP model}

The model employed in this work is the many-body SNAP force field~\cite{Thompson2015316,Lunghi2019}, whose main
features are briefly outlined here. In general we can write the total energy as 
\begin{align}\label{Etot}
    E_\mathrm{total}(\vv{r}^N)=E_\mathrm{ZBL}(\vv{r}^N)+E_\mathrm{SNAP}(\vv{r}^N)\:,
\end{align}
where $\vv{r}^N$ is the configuration vector containing the $N$ atomic positions. The Ziegler-Biersack-Littmark (ZBL)~\cite{ZBL1985} 
empirical potential, $E_\mathrm{ZBL}(\vv{r}^N)$, is introduced to provide a repulsive short-range interaction, while 
$E_\mathrm{SNAP}$ captures the effects arising from the local atomic environments. SNAP assumes that the energy 
of a collection of atoms can be decomposed into separate atomic contributions, $E^i_\mathrm{SNAP}$. The energy associated 
to the $i$-th atom, $E^i_\mathrm{SNAP}$, then depends on the local atomic environment, which is described by using 
the bispectrum components~\cite{GAP2010} ($\vv{B}^i$= {$B^i_1$, ...,$B^i_K$}). Thus the SNAP potential writes as a linear 
combination of $\vv{B}^i$, namely as
\begin{align}\label{Esnap}
    E_\mathrm{SNAP}=\sum_{i=1}^NE_\mathrm{SNAP}^{i}(\vv{B}^i)=
  {\sum_{i=1}^{N}\{\beta_0^{\alpha_i}+\sum_{k=1}^{K}\beta_k^{\alpha_i}B_k^i\}}\:.
\end{align}
Here $\alpha_i$ describes the chemical identity of atom $i$, while $\beta^\alpha_k$ are the coefficients of expansion for
the atoms of type $\alpha$. $K$ is the number of bispectrum components used to describe the local atomic environment. 
This is related to the maximum angular momentum taken in the definition of the hyper-spherical harmonics expansion
(see next paragraph).

In practice SNAP knows about the local atomic environment of each atom by projecting the local neighbour density within 
a radial cutoff, $R_\mathrm{cut}$, onto a basis of hyper-spherical harmonics in four dimensions. This choice of representation 
for the atomic positions is invariant under rotation, atomic permutation and translation. The details of how the ${B}^i_k$ are
calculated are provided in reference~\cite{GAP2010}. In general, specifying small values for $R_\mathrm{cut}$ and $K$ produces
a coarse description of the local atomic environment. In contrast, large values will enable one to resolve fine structural
differences. The coefficients $\beta^\alpha_k$ are optimized to reproduce the energies of a large set of configurations
computed with density functional theory (DFT). The training process is performed by minimizing the penalty function, 
\begin{align}\label{Emin}
\Delta E  =  \lvert\omega \cdot E_\mathrm{FF}-E_\mathrm{DFT}\rvert^2+\lambda\lvert\beta\rvert^2\:,
\end{align}
where $E_\mathrm{FF}$ and $E_\mathrm{DFT}$ are, respectively, the energy predicted by Eq.~(\ref{Etot}) and the 
corresponding DFT reference. The second term of Eq.~(\ref{Emin}) penalizes large $\beta^\alpha_k$ values in
order to avoid over-fitting problems, with $\lambda$ controlling this regularization process. The weight of the parameter 
$\omega$ is designed to control the relative relevance of different configurations. In this work, all configurations
are set equivalent, so that $\omega$=1. Detailed convergence tests for  $\lambda$, $K$ and $R_\mathrm{cut}$, 
are shown in Fig.~S1 of the supplementary information (SI). Their final values, together with the ZBL potential 
parameters are summarized in Table~\ref{table1}. 
\begin{table}[ht]
\centering
\caption{SNAP and ZBL potential parameters used for our optimized ZrB$_2$ force field.  
$\lambda$ is the regularization parameter; $K$ the number of bispectrum components; $R_\mathrm{cut}$ 
the cutoff radius of the neighbor density function; $R_{zbl, i}$ and $R_{zbl, o}$ the inner and outer cutoff 
of the switching function for ramping the energies, forces and curvatures smoothly to zero; $Z_\mathrm{Zr}$ and 
$Z_\mathrm{B}$ the atomic number of the Zr and B atoms.
  }
\begin{tabular}{l|l|l}
\hline\hline
Potential & Parameter & Value \\ \hline
SNAP & $\lambda$         &  0        \\
SNAP & $K$               &  55        \\
SNAP & $R_\mathrm{cut}$  &  4.5 \AA     \\
ZBL & $R_{zbl, i}$       &  4.0 \AA   \\
ZBL & $R_{zbl, o}$       &  4.8 \AA   \\
ZBL & $Z_\mathrm{Zr}$    &  40        \\
ZBL & $Z_\mathrm{B}$     &  5        \\
\hline\hline
\end{tabular}
\label{table1}
\end{table}

\subsection{DFT datasets}
A pool of diverse atomic configurations needs to be included in the DFT reference dataset in order to represent the 
widest variety of local atomic environments. Such diversity is critical for the construction of a robust force field able 
to fully capture the features of the potential energy surface. Our training dataset comprises a total of approximately 
150,000 local atomic environments extracted from 1850 configurations, while the test dataset contains around 
137,000 local atomic environments. Those configurations include randomly distorted supercells (1109), deformed 
cells (1033), molecular-dynamics snapshots (1457), etc. A detailed description of our DFT dataset and its generation 
scheme is tabulated in Table~\ref{table2}. Notably different classes of data sample different regions of the potential energy 
surface. In particular each class is designed to target at a variety of properties in the following way (see Table~\ref{table2} for 
the details of each class):
\begin{itemize}
   \item {\it EOS I/II} map the elasticity and the general profile of the potential energy surface;
   \item {\it Distorted  I/II} capture the phonon band-structure and the thermal conductivity;
   \item {\it Deformed I/II/III} describe the $T$-dependent elasticity, thermal expansion and anharmonicity 
   at high temperature;
   \item {\it MD snapshots I/II} ensure the dynamic stability of the crystalline structure.
\end{itemize}

\vspace*{1cm}
\begin{table*}[ht]
\small
\centering
\caption{Generation protocol for the DFT reference configurations. For each data class we list the number of inequivalent
configurations computed, $N_\mathrm{conf}$, the total number of atoms included in the simulation cell, $N_\mathrm{atoms}$, 
and the range of volume expansion/compression, $V / V_0$, measured with respect to the $T=0$~K equilibrium volume, $V_0$.
The last column describes the procedure used to generate the specific dataset. The {\it EOS} (equation of state) class includes 
configurations obtained by altering the $a$ and $c$ lattice parameters separately, while keeping the atomic positions frozen at 
their equilibrium values. 
In the {\it Distorted} class atoms are randomly displaced from their equilibrium positions. The {\it MD snapshots} class
includes configurations taken from molecular dynamics (MD) simulations in the $NVT$ ensemble performed with the on-the-fly SNAP 
force field in the temperature range 100~K-1800~K. The {\it Deformed} class contains configurations obtained for supercells
strained along one of the elementary deformation modes (see details in the SI), while the atomic positions are either from
MD simulations or are obtained by random atomic displacement.}
\resizebox{\textwidth}{!}{\begin{tabular}{lcccl}
\hline\hline
Data Class      &  $N_\mathrm{conf}$   &  $N_\mathrm{atoms}$  &  $V / V_0$ (\%)   &     Generation scheme \\
\hline
{\it EOS I}          & 3 & 3 &  1.5 -- 3.0  &   $a$ and $c$ are altered with the atoms kept at their equilibrium positions \\
{\it Distorted I}     &  309  & 81  &  0.0 -- 3.7 & The Zr and B atomic positions are randomly displaced from equilibrium  \\
                      &    &  &  &           with a maximum displacement of $d_\mathrm{max}$~=~0.01/0.05/0.08/0.10/0.15 \AA   \\
{\it Distorted II}    &  800 & 81  &  1.7 -- 3.7  & Only the B atomic positions are randomly displaced from equilibrium \\
                      &   &   &    &      with a maximum displacement of $d_\mathrm{max}$~=~0.05/0.07/0.13/0.20~\AA    \\
{\it MD snapshots I}  &   338  & 81  &  0.6 -- 2.0 & MD snapshots collected every 0.1 ps from an $NVT$ ensemble \\
{\it Deformed I}      &   400  & 81  & 1.4 -- 4.0  & The atomic positions are from MD snapshots (0.1~ps in the $NVT$ ensemble), \\
                     &    &         &            & while the cell is strained by 1\% along various deformation directions   \\
\hline
{\it MD snapshots II} &   1119      & 81      & -1.3 -- 4.1 &    MD snapshots collected every 0.1 ps from an $NVT$ ensemble \\
{\it Deformed II}    &   120  & 108 &  1.0    &  The atomic positions are randomly displaced with $d_\mathrm{max}$~=0.10~\AA ,\\ 
                     &        &    &      &   while the  cell is strained by 1\% along various deformation directions  \\
{\it Deformed III}   &   513  & 81 &   -3 -- 2.0 &  The atomic positions are randomly displaced with $d_\mathrm{max}$~=0.13~\AA, \\
                     &        &   &       &   while the cell is strained by 1\% along various deformation directions \\
\hline
{\it EOS II}   &   242   & 3  &  -28.5 -- 40.6 &   $a$ and $c$ are altered with the atoms kept at their equilibrium positions\\
 \hline\hline
\end{tabular}}
\label{table2}
\end{table*}

The energies and the atomic forces, as calculated from DFT for the aforementioned dataset, are all collected. The SNAP force
field for ZrB$_2$ is then trained on the energies of the data classes: {\it EOS I},  {\it Distorted  I\&II}, {\it Deformed I}
and {\it MD snapshots I}, while {\it EOS I}, {\it Deformed   II \& III} and {\it MD snapshots II} are used for testing and
validation purposes. The diversity of our training and test sets is further illustrated by the dimension reduction analysis 
presented in Fig.~S2 of the SI, while their energy distribution is presented in Fig.~S3.

\subsection{Computational details}
All DFT calculations are performed using the plane-wave basis projector augmented wave method~\cite{Blochl1994} as implemented
in the VASP code~\cite{Kresse1999}.  The generalized gradient approximation of the exchange and correlation functional with the
Perdew-Burke-Ernzerhof (PBE) parametrization~\cite{Perdew1996} is used throughout. In addition, the damped van der Waals
correction, DFT-D2, is included to approximate dispersive interactions~\cite{Grimme.2006}. The reliability of the PBE+D2 method 
in describing transition metal diborides has been validated already in reference~\cite{Zhang2018}. The Brillouin zone is sampled
by using the Monkhorst-Pack method with a \textit{k}-point spacing of 0.02~\AA$^{-1}$. The kinetic energy cutoff of the plane
waves is set to 500~eV. Those parameters have been tested to ensure an energy convergence of 1~meV/atom. 

All molecular dynamic (MD) simulations are performed with the LAMMPS package~\cite{PLIMPTON19951} with a time step 
of 1~fs. The damping parameters for the temperature and pressure are 100 and 800 fs, respectively. The calculations of the 
thermal expansion, thermal conductivity and elastic moduli are carried out using $6\times 6 \times 6$ supercells. More details 
on the MD simulations can be found in the SI.

\subsection{Accuracy of SNAP force field}
\begin{figure}[H]
\centering
\includegraphics[width=2.8in]{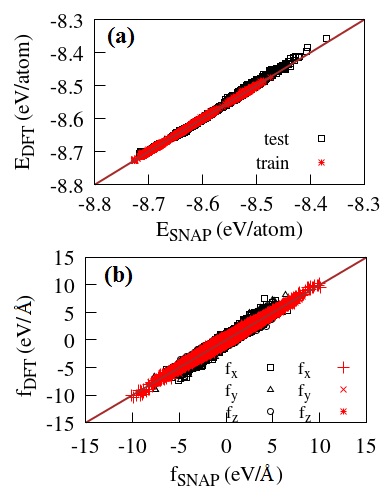}
\caption{(color online) SNAP predictions against the DFT references for (a) the energies (eV/atom) and (b) the atomic forces
(eV/\AA). The training (test) set contains 1850 (1752) configurations, with roughly 449,000 (411,000) force data. Note that the
forces were not explicitly included in the training.}
\label{fig.1}
\end{figure}

The SNAP-predicted energies (in unit of eV/atom) and atomic forces (in unit of eV/\AA) compare extremely well to their DFT 
references. The linear regression analysis of energies in Fig.~\ref{fig.1}(a) gives the angular coefficients of 1.0004 and
1.0006 for the training and test sets, respectively. Being so close to 1 demonstrates that the quality of the SNAP energies is
almost the same as that of DFT, regardless on whether the configurations were included or not in the training process.  
At the same time [see Fig.~\ref{fig.1}(b)] the linear regression coefficients computed over the training (test) set for the
atomic forces along the $x$, $y$ and $z$ directions are 0.9878 (0.9643), 0.9869 (0.9611) and 1.0099 (0.9932), respectively. 
This quality translates in a root mean squared error (RMS) over the energy of 1.3~ meV/atom and 4.7~meV/atom, respectively 
for the training and test set. The RMS for the atomic forces is 25~meV/\AA\ (45~meV/\AA) for the training (test) set. In order 
to put these results in prospective, a recent high-quality gaussian approximation potential for carbon-based materials has 
achieved a RMS of 28~meV/\AA\ and 270~meV/\AA\ over the in-plane forces of graphene and amorphous carbon, 
respectively~\cite{Rowe2018}. As such, our generated SNAP appears to yield a near to quantum-chemistry accuracy and it
is fully competitive with other advanced force field classes.

\begin{figure}[H]
\centering
\includegraphics[width=2.8in]{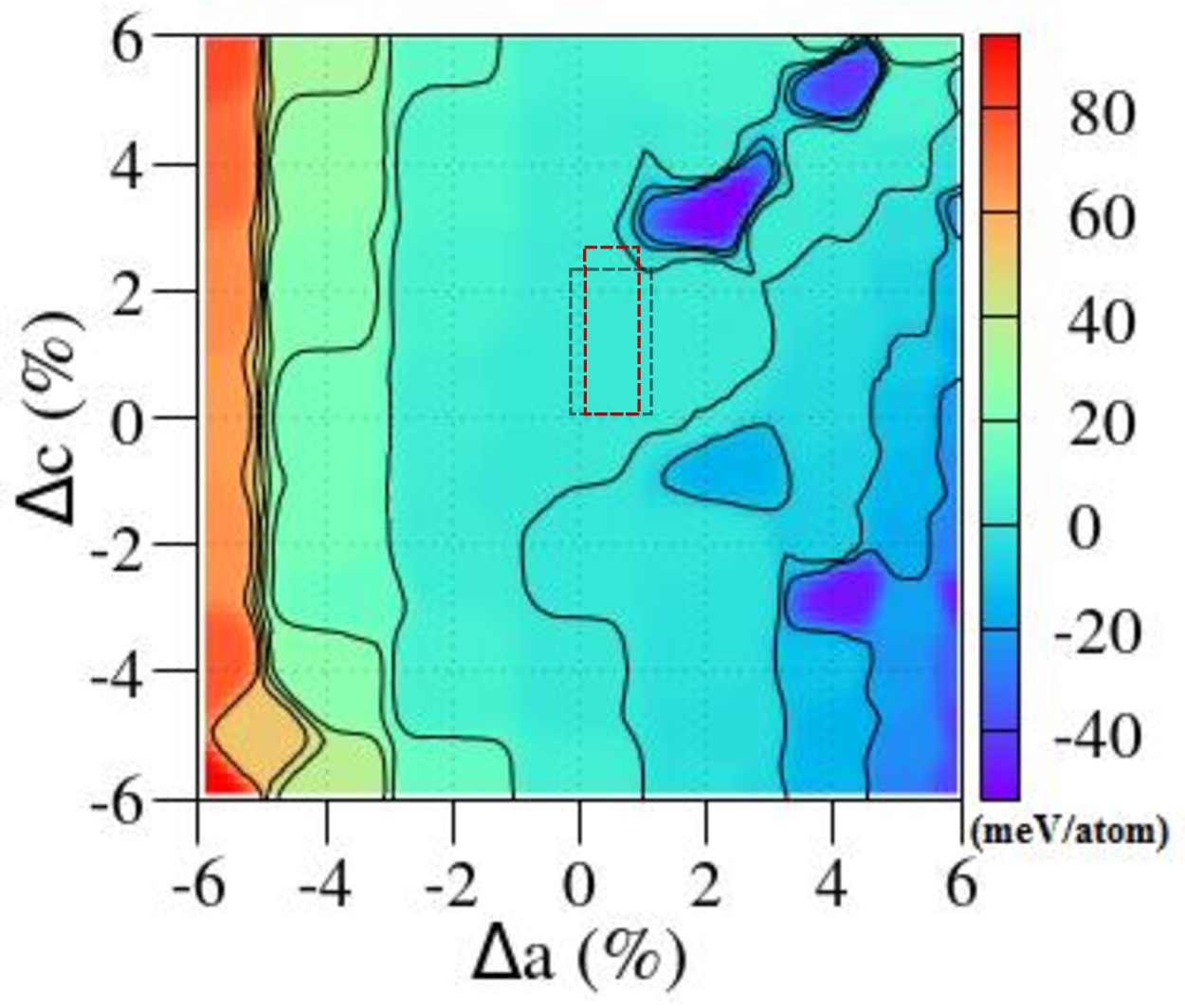}
\caption{(color online) Evaluation of the SNAP accuracy in computing the total energy as a function of the variation 
of the lattice parameters $\Delta a$ and $\Delta c$ (in \%). The color scale indicates the RMS energy error (in meV/atom).
The grey and red boxes show the relatively small region included in training process and the relevant thermal expansion domain
upto 2500 K, respectively.}
\label{fig.2}
\end{figure}

Furthermore, the SNAP force field demonstrates an excellent performance on describing the energy landscape against
the variation of structural parameters as shown in Fig.~\ref{fig.2}. In this case the target property is the dependence of the
total energy over the lattice parameters, $a$ and $c$, which are changed with respect to their equilibrium values. The DFT
references for this test are contained in {\it EOS II} dataset, which is not used for the training. Note, in fact, that the
largest lattice strain included in the training set is of only 1.3\% ({\it EOS I} dataset, see the corresponding region marked
using the grey box in Fig.~\ref{fig.2}). In Fig.~\ref{fig.2} it is easy to observe that the SNAP predictions are within a
deviation of 20~meV/atom from their corresponding DFT references for the lattice deformation region of -2.0\% $\leq \Delta a
\leq 2.0\%$ and -6.0\% $\leq \Delta c \leq$ 2.0\%. A relatively larger deviation shows up only for rather severe deformations,
namely $\Delta a$ beyond $\pm$5.0\%.  Since ZrB$_2$ is an extremely stiff crystal (Young's modulus reported between 490~GPa and 
526~GPa~\cite{Okamoto2010,Nakamori2015}) with relatively small thermal expansion (see the relevant region noted within the red
box of Fig.~\ref{fig.2}), the feasible thermal deformation region is limited and is well covered within the high-accuracy 
domain of SNAP force field.

\begin{figure}[H]
\centering
\includegraphics[width=2.8in]{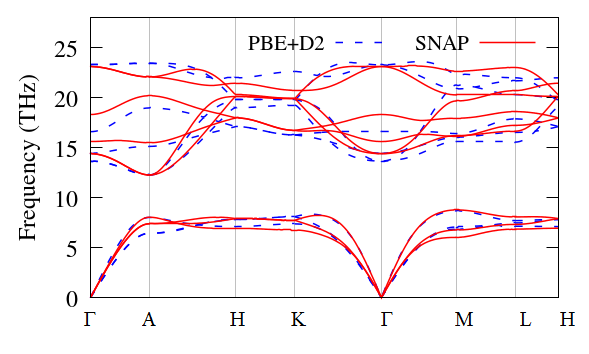}
\caption{(color online) Phonon dispersion curves calculated with SNAP (in red) and DFT PBE+D2 (in blue). }
\label{fig.3}
\end{figure}

Finally, Fig.~\ref{fig.3} illustrates the excellent performance of the SNAP force field in calculating the phonon bands as
compared with the DFT results. In both cases the dynamical matrices are derived using the finite difference method, 
for which the $3\times3\times3$ supercells with finite atomic displacement of 0.01 \AA~are adopted. Then, a Fourier
interpolation over a dense $108\times108\times108$ mesh, performed with the {\sc phonopy} code~\cite{Togo2015,TOGO20151}, is
used to obtain a smooth frequency dispersion. The good agreement of the phonon bands computed with SNAP and DFT in
Fig.~\ref{fig.3}, indicates that our SNAP force field is not only capable of describing the potential energy surface, but also
to qualitatively account for its derivatives.

\section{Applications}
In what follows we will evaluate the ability of our force field to predict several $T$-dependent fundamental physical 
properties of ZrB$_2$. We will begin with its performance around room temperature, then move to the high-temperature
region. Finally we will discuss the challenges encountered at extremely high temperatures (above 2000~$\degree$C).

\subsection{Room-temperature properties}
We start our analysis by looking at the ZrB$_2$ structural and elastic properties near room temperature. In Table~\ref{table3}
we report the calculated values for the lattice parameters, $a$ and $c$, the elastic constants, $c_{ij}$, the bulk modulus, 
$B$, the Young's modulus, $E$, and the shear modulus, $G$. These are compared to the measured experimental 
values~\cite{Gasparov2001,Okamoto2010,Nakamori2015} and to our DFT PBE+D2 predictions. Furthermore, we include in 
the table the published data obtained with two different parameterizations of the Tersoff potential~\cite{Daw2011}. 

\vspace*{1cm}
\begin{table*}[ht]
\small
\centering
\caption{Room-temperature structural and elastic properties of ZrB$_2$ calculated with our SNAP force field. Results are
compared to experiments, to DFT (PBE+D2) estimates and to predictions obtained with Tersoff potentials.~\cite{Daw2011}}
\begin{tabular}{c|c|c|c|c|c}
\hline
\hline
               &  Exp.~\cite{Gasparov2001,Okamoto2010,Nakamori2015}  &   PBE+D2~\cite{Zhang2018}    & Tersoff$_{P1}$~\cite{Daw2011}    & 
               Tersoff$_{P2}$~\cite{Daw2011}   & SNAP    \\
\hline
$a$ (\AA)        &  3.170                      &   3.162     & 3.140         & 3.131       & 3.163  \\
$c$ (\AA)        &  3.532                      &   3.493     & 3.547         & 3.484       & 3.486  \\
$c_{11}$ (GPa)  &  568                        &   584       & 422           & 575         & 580    \\ 
$c_{33}$ (GPa)  &  441                        &   431       & 320           & 817         & 449    \\
$c_{44}$ (GPa)  &  258                        &   253       & 119           & 246         & 336    \\   
$c_{12}$ (GPa)  &  57                         &   56        & 156           & 213         & 63     \\
$c_{13}$ (GPa)  &  121                        &   116       & 171           & 301         & 104    \\
$B$ (GPa)        &  224-240                    &   251       & 240           & 400         & 238    \\
$G$ (GPa)        &  216-232                    &   208       & 118           & 211         & 268    \\
$E$ (GPa)        &  490-526                    &   488       &    --          &   --         & 586    \\
\hline
\end{tabular}
\label{table3}
\end{table*}
In general we find a rather good agreement over the entire range of properties between our SNAP predictions and the DFT target.
Importantly, since DFT at the level of PBE+D2 provides an excellent description of the experimental data, such agreement with
experiments is transferred to our SNAP force field. Going into more detail, we note that DFT~\cite{Zhang2018} slightly
underestimates the experimental lattice parameters~\cite{Gasparov2001}, being $a=3.162$~\AA~and $c=3.493$~\AA\ instead of
$a=3.170$~\AA~and $c=3.532$~\AA. SNAP, which is trained over DFT PBE+D2 data, as expected maintains such underestimation and
returns us the lattice parameters $a=3.164$~\AA~ and $c=3.486$~\AA. Notably, SNAP outperforms the two available Tersoff models,
which further overbind, in particular along the $a$ axis (3.131~\AA~$<a<$~3.140~\AA). \\

Turning our attention to the elastic constants and moduli we remark again the good agreement between DFT and experiments
~\cite{Okamoto2010,Nakamori2015} and the ability of SNAP to reproduce the DFT results, and hence the experimental ones.
The only exception is for the $c_{44}$ constant, that SNAP overestimates by about 25\%. Such error propagates in the
overestimation of both $G$ and $B$. Note that we could have improved the SNAP description of the elastic properties by including
in the training set additional configurations describing low temperature distortions. However, our objective is to obtain a
potential performing over a wide temperature range, so that our training set needs to remain well balanced. As such, we
effectively trade some accuracy in describing the room-temperature elastic tensor for a large dynamic range. Finally, as for 
the structural properties, also for the elastic ones SNAP outperforms completely the available Tersoff 
potentials~\cite{Daw2011}, which in general are not capable to even rank the magnitude of the various elastic constants. 

\subsection{High-temperature properties}
\begin{figure}[H]
\centering
\includegraphics[width=2.8in]{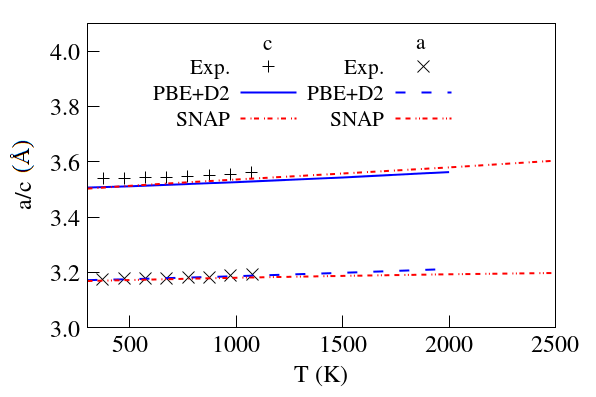}
\caption{(color online) Variation of the lattice parameters, $a$ and $c$ (in \AA), with temperature. The SNAP results 
(in red) obtained from $NPT$ molecular dynamics (averaged over 100~ps) are compared with the those from DFT 
PBE+D2 (in blue) and experiments~\cite{Okamoto2010} (in black). }
\label{fig.4}
\end{figure}

We now move to explore how our optimized SNAP force field is able to describe the various properties in a temperature 
range comprised between 300~K and 2500~K. Firstly, we investigate the thermal expansion effects by calculating the 
temperature-dependent lattice parameters, $a(T)$ and $c(T)$. This is performed by sampling the $NPT$ molecular-dynamics 
trajectories over 100~ps for various temperatures. The SNAP potential predicts the thermal expansion along the $c$ axis 
to be relatively higher than that along $a$, as shown in Fig.~\ref{fig.4}. Experimental data~\cite{Okamoto2010} are available 
up to 1,200~K, a temperature range described extremely well by our SNAP force field, which again traces well the DFT
results obtained using the quasi-harmonic approximation of phonon calculations at the PBE+D2 level. The excellent agreement
between SNAP and DFT, in fact, extends up to 2,000~K. Note the latter is a static calculation method, being computationally
heavy and the anharmonic effects can only be suitable considered within finite temperatures, below 2000 K.
Most importantly the SNAP predictions can reach 2,500~K, meaning that our atomic potential is capable to explore the 
ultra-high temperature region, where experiments are difficult to perform and are not yet available. 

\begin{figure}[ht]
\centering
\includegraphics[width=3.3in]{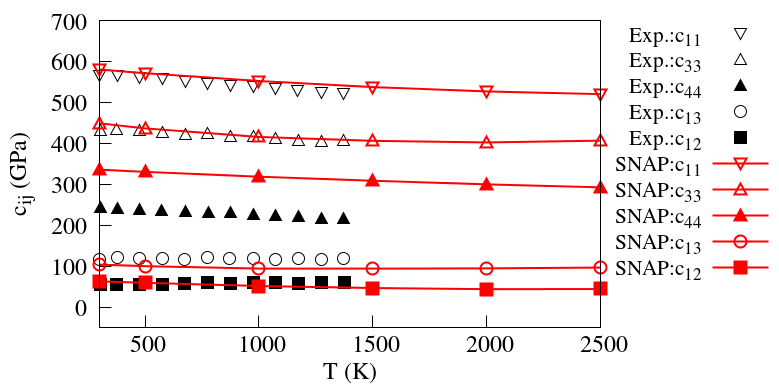}
\caption{(color online) Variation of the elastic constants, $c_{ij}$ (GPa), with temperature. The SNAP 
results (in red) are extracted from the $NVE$ trajectories over a 250~ps interval after an initial equilibration 
of 150~ps. Results are compared with available experiments~\cite{Okamoto2010} (in black). }
\label{fig.5}
\end{figure}

We then move to determine the temperature dependence of the elastic tensor, $c_{ij}$, as computed from SNAP. This quantity 
is determined from the trajectories of $NVE$ molecular dynamics simulations by using the explicit deformation
method~\cite{Clavier2017} (see computational details in the SI). During the simulations the lattice parameters as a function
of temperature, $a(T)$ and $c(T)$, are fixed to those previously derived by SNAP and reported in Fig.~\ref{fig.5}. 

The results of $c_{ij}$($T$) are presented in Fig.~\ref{fig.5}, where a good agreement with experiments~\cite{Okamoto2010}
is achieved over the temperature range where experiments are available. The agreement is particularly good for 
the elastic tensor components $c_{11}$, $c_{33}$, $c_{12}$ and $c_{13}$, while that for $c_{44}$ is less
satisfactory. Note that, in general, the deviation of $c_{44}$ with respect to the experimental value at high
temperature simply follows the same deviation found at room temperature (see Table~\ref{table3}). It is important 
to point out that $c_{44}$ is related to the elastic response of the cell to shear strains. Our training data set 
is dominated by configurations constructed over undistorted cells (1447), and only a small fraction contains 
biaxial deformed ones (403). In principle, further improvements in $c_{44}$ can then be obtained by increasing the 
training set to add sheared configurations. This, however, should be done without unbalancing 
the training set, that otherwise will be unsuitable to capture the behaviour of the remaining $c_{ij}$ constants. 
Considering the general good agreement of the overall elastic tensor this further refinement has not been 
performed. 

Finally we turn our attention to the variation of the thermal conductivity with a wide range of temperatures, which is
satisfactorily derived using our SNAP force field. In this case the lattice thermal conductivity tensor, $\kappa_{ij}$, is
calculated by using the equilibrium Green-Kubo formalism~\cite{Zwanzig1965}. In particular, $\kappa_{ij}$ is evaluated from
$NVE$ molecular dynamics simulations of heat flux up to 40~ns, a time needed to achieve convergence at temperatures 
comprised between 300~K and 2500~K (see computational details in SI).  
\begin{figure}[H]
\centering
\includegraphics[width=2.8in]{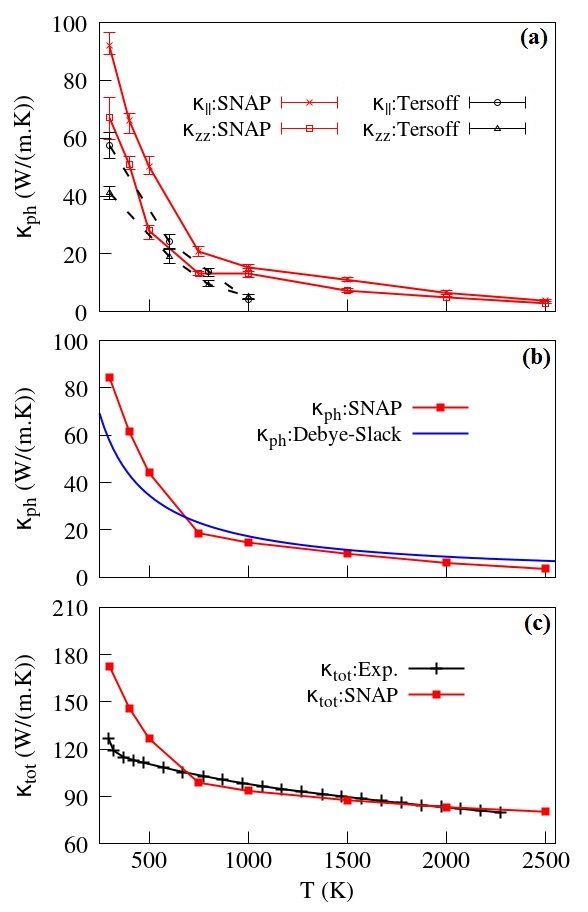}
\caption{(color online) Thermal conductivity, $\kappa$ (in units of W/m$\cdot$K), as a function of the temperature. 
In panel (a) the SNAP predictions (in red) for the in-plane, $\kappa_{||}$, and out-of-plane, $\kappa_{zz}$, components
of the lattice thermal conductivity are compared with results obtained with a Tersoff potential (in black)~\cite{Lawson2011}.
Panel (b) shows the orientation-averaged lattice thermal conductivity computed with SNAP (in red) together with an estimate 
obtained from the Debye-Slack model (in blue). Panel (c) shows the the total calculated thermal conductivity, 
$\kappa_\mathrm{tot}=\kappa_\mathrm{ph}+\kappa_\mathrm{el}$ (red), in comparison with experiments (black)~\cite{Okamoto2010}.
In this case the electronic thermal conductivity is estimated based on the Wiedmann-Franz law.}
\label{fig.6}
\end{figure}

The calculated in-plane, $\kappa_{||}$, and out-of-plane, $\kappa_{zz}$, lattice thermal conductivities as a function of
temperature are shown in Fig.~\ref{fig.6}(a). SNAP returns us an anisotropic lattice thermal conductivity for temperatures up to
about 1,000~K, while it is essentially isotropic {at elevated temperatures}. The anisotropy remains, however, 
rather moderate even in the low temperature region. Available theoretical results obtained with a Tersoff force field are also
presented~\cite{Lawson2011}. These are limited to temperatures up to 1,000~K and constantly return $\kappa$ values smaller than
those computed by SNAP, although also the Tersoff potential captures the anisotropy. It was argued in the past~\cite{Lawson2011} 
that a phonon thermal conductivity of  5~W/m$\cdot$K at 1000~K, as obtained with the Tersoff force field, is too low 
when compared to experiments. SNAP improves on such determination and predicts $\kappa$ to be around 
15~W/m$\cdot$K at 1000 K. This improved consistency with experiments originates from the accurate description of 
the interatomic interaction offered by SNAP. Since the lattice thermal conductivity is dominated by phonon scattering 
processes, the accurate SNAP description of the forces makes it possible to accurate calculate the heat flux and thereby 
to successfully predict $\kappa$. This is true not only at relatively lower temperature, but over a wide temperature range 
extending up to ultra-high temperatures, as we will demonstrate in the following.

The lattice thermal conductivity at high temperatures can be estimated by using the Debye-Slack 
model~\cite{Slack2006,Zhang2016b}. In Fig.~\ref{fig.6}(b) we present the average lattice thermal 
conductivity, $\kappa_\mathrm{ph}$, in comparison with the Debye-Slack prediction (see details in SI). 
The $\kappa_\mathrm{ph}(T)$ curve is defined as the square average of the diagonal components of 
the conductivity tensor, $\kappa_\mathrm{ph}=1/3\sqrt{\kappa_{11}^2+\kappa_{22}^2+\kappa_{33}^2}$. 
We find that above 1000~K, our SNAP results agree well with the $\kappa_\mathrm{ph}$ determined 
with the Debye-Slack model. At lower temperatures the agreement is less good with the SNAP predictions 
being constantly above the Debye-Slack curve. This is expected, since the Debye-Slack model is designed 
to estimate the thermal resistivity at temperatures above the Debye temperature, $\theta_\mathrm{D}$. 
In fact, it assumes Umklapp scattering to be the dominant process limiting the thermal conductivity and 
completely ignores contributions from scattering to the optical phonon branches. This kind of approximation 
works in the high temperature limit, but underestimates $\kappa_\mathrm{ph}$ for $T<\theta_\mathrm{D}$. 
In contrast, our SNAP force field provides a robust tool to perform molecular dynamic simulations of heat 
transfer over the full temperature spectrum. At the same time it bears a much higher accuracy and efficiency.

Finally in Fig.~\ref{fig.6}(c) we compare the total thermal conductivity, $\kappa_\mathrm{tot}=\kappa_\mathrm{ph}+\kappa_\mathrm{el}$, 
with experiments. Here the $\kappa_\mathrm{tot}(T)$ curve is obtained by adding to the SNAP-calculated 
lattice thermal conductivity, the electronic contribution, $\kappa_\mathrm{el}$. The latter is estimated by using the
Wiedemann-Franz law with data for the temperature-dependent electrical conductivity taken from
available experiments (see SI). Notably, our $\kappa_\mathrm{tot}$ is in excellent agreement with that measured in 
experiments~\cite{Okamoto2010} for all temperature above 600~K. Deviations at low temperature are possibly due 
to phonon scattering to crystal defects, grain boundaries and surfaces. These mechanisms, which in general reduce 
the lattice conductivity, are not included in our SNAP calculations. Note that reductions in the electrical conductivity 
due to defect scattering are already taken into account by the use of experimental data when determining $\kappa_\mathrm{el}$.

\subsection{Challenges emerging at ultra-high temperatures}
The SNAP force field that we have just presented (named `SNAP-I') describes remarkably well several 
temperature-dependent physical properties of ZrB$_2$ up to 2500~K. At the same time `SNAP-I' maintains 
the dynamical stability of ZrB$_2$ over all the temperatures investigated. The next question is how well can 
the SNAP force field simultaneously describe the dynamic stability of ZrB$_2$ above 2600~K and maintain 
high accuracy in determining the $T$-dependent physical properties. The construction of a force filed describing 
the ultra-high temperature region requires additional configurations to be included in the training set. These should
explore geometries characterized by larger bond lengths and lattice distortions. We have performed such exercise 
and generated a full range of novel SNAP force fields.

\begin{figure*}[ht]
\centering
\includegraphics[width=5.6in]{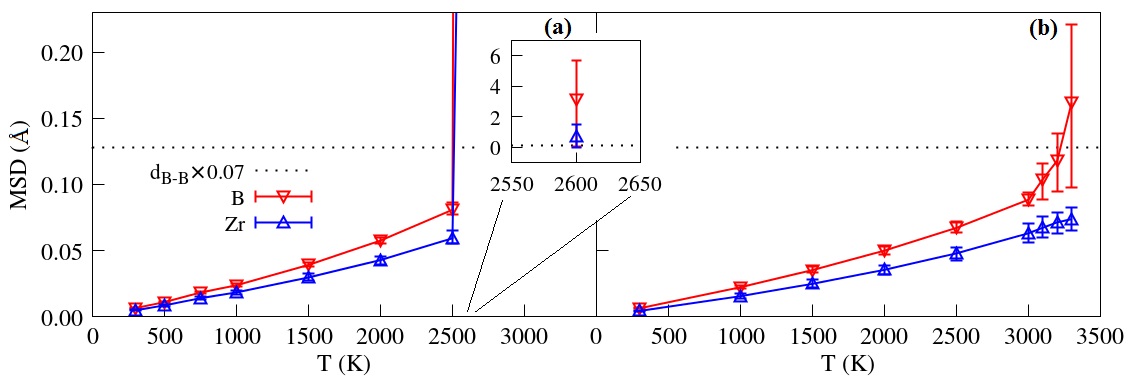}
\caption{(color online) Temperature-dependent mean-squared displacement (MSD, in \AA) for Zr (in blue) and B  (in red) atoms
predicted using (a)  `SNAP-I' and (b) `SNAP-II', respectively. The inset shows the behaviour of `SNAP-I' at 2,600~K. The dashed
horizontal line marks the boundary established by Lindermann's law for melting, MSD~$>0.07\times d_\mathrm{B-B} \sim0.128$~\AA,
with $d_\mathrm{B-B}$ being the B-B equilibrium bond length.}
\label{fig.7}
\end{figure*}

Our results suggest that a SNAP force field able to retain the dynamic stability of ZrB$_2$ to temperatures up to 3,300~K
and possibly above can be constructed. However, in the construction we have to relax its performance in predicting the
elastic and mechanical properties at low temperature. In all cases, unfortunately, an extension of the dynamical range 
of the force field above 2,600~K always comes at the expenses of the accuracy in describing the mechanical properties 
below 2,000~K. In particular we were not able to match the description of the temperature dependent elastic tensor provided 
by `SNAP-I'. In any case, the results for our best attempt, `SNAP-II', are presented in Fig.~\ref{fig.7} to demonstrate the 
feasibility of extending the dynamic stability. 

In the Fig.~\ref{fig.7}, panel (a) and (b) present the mean squared displacements (MSDs) computed with SNAP-I and SNAP-II, 
respectively, over an $NVT$-ensemble trajectory of 1~ns. As expected the MSD of the lighter B atoms is consistently 
above that of Zr at any temperature, regardless of the SNAP version used. The MSD increases with temperature linearly 
below 1500 K, which is similar for the two force fields. However, at 2,600~K the MSDs calculated for SNAP-I suddenly 
take extremely large values, indicating that the force field is no longer able to capture the movement of certain atoms. 
In contrast, SNAP-II does not present such discontinuity in the MSDs, which retain a smooth dependence with the 
temperature. More significant thermal disturbance is observed for the B atoms at temperatures between 3,200~K and 
3,300~K. It is, in fact, around these values that the MSD for B surpasses the boundary established by Lindermann's law 
for melting, i.e. $0.07\times d_\mathrm{B-B} \sim0.128$~\AA, where $d_\mathrm{B-B}$ is the equilibrium B-B bond length. 
We then conclude that, in general, SNAP is capable of describing the dynamics at ultra-high temperatures, although 
with a compromise about its low temperature accuracy. Future work is needed to develop a consistently accurate 
force field capable to describe ZrB$_2$ up to the melting point.

\section{Conclusions}
In this work we have developed a machine learned SNAP force field for ZrB$_2$, one important ultra-high temperature 
ceramic for the fabrication of the hot components in the aerospace and nuclear industries. After having detailed the
procedure to construct a balanced training set, we have shown how such force field can describe the structural, mechanical,
thermodynamics, and thermal transport properties, both at room and at high temperatures. In particular, it provides a reliable 
and efficient tool to compute the lattice thermal conductivity across a wide temperature range. Relying on the
richness of its training dataset and the limited number of model parameters our SNAP can successfully and efficiently
capture the $T$-dependent properties from room temperature to 2500~K. This covers completely the temperature range required by
applications, making our SNAP potential an important tool for exploring materials properties not easily accessible by experiments.
Overall our work demonstrates the possibility to construct DFT-accurate force fields for materials operating in extreme conditions.

\section*{Acknowledgement} 
We acknowledge G.C. Sosso (University of Warick) for the useful discussions. This work is supported by the European Union's
Horizon 2020 ``Research and innovation programme'' under the grant agreement No.685594 (C$^3$HARME). Computational resources
have been provided by the Irish Center for High-End Computing (ICHEC) and the Trinity Centre for High Performance Computing
(TCHPC). The code for fitting the SNAP force field can be found at https://github.com/lunghiale/fitsnap.

\section*{References}


\end{document}


\maketitle
\singlespacing

\section{Computational details}
\subsection{Convergence tests}

The hyper-parameters of the SNAP force field (FF) are set after several convergence tests, which are illustrated in 
Figs.~S1(a-c). Fig.~S1(a) shows how the energy accuracy (RMS) varies with the radial cutoff distance, $R_\mathrm{cut}$. 
With increasing $R_\mathrm{cut}$, first the RMS decreases and shows a minimum for $R_\mathrm{cut}$ in the  
4.0~\AA~$<R_\mathrm{cut}<$~6.0 \AA~range. However, the RMS increases again for $R_\mathrm{cut}$ beyond 7~\AA. 
This holds true for both the training and the testing sets. The large error for $R_\mathrm{cut}$ below 3.0~\AA\ is 
easy to understand, since only the 1$^{st}$ and/or 2$^{nd}$ nearest neighbors are considered as a part of the local atomic 
environments defining $E^i_\mathrm{SNAP}$. At the same time it is not possible to increase $R_{cut}$ indefinitely in order to lower
down RMS. In fact, its upper bound is limited by the computational costs and by the size of the supercells used to construct the 
DFT datasets. Here, we adopt a $R_{cut}$ value of 4.5 \AA. This is around half of the lattice parameters of the DFT reference
structures ($a=$~9.516, $c=$~10.628). Such value is large enough to include as far as the 5$^{th}$ coordination shell of B atoms. 

Figure S1(b) illustrates how the accuracy over the energy (RMS) improves with the number of bispectrum components, $K$, 
namely with extending the descriptors space. As shown in the figure, a significant reduction of the RMS appears for $K>$~30.  
Thereafter the improvement becomes rather moderate upon further increasing $K$. The RMS converges at $K$= 55 for both the 
training and testing sets. Since using a huge number of parameters will jeopardise the efficiency of the SNAP and will also 
likely generate over-fitting, $K$=55  is chosen in this work. 

In order to check whether there is an overfitting problem, we have carried out a Tikhonov regularization analysis by monitoring the 
RMS change with the regularization parameter, $\lambda$. This is performed by using the {\it MD snapshots} and {\it Deformed} training 
sets. Normally a large $\lambda$ value can help in suppressing the overfitting problem. If overfitting is present, the RMS over the training 
set will be very low, but that over the testing sets will surge. This is for sure not the case here, as seen in Fig. S1(c). A very low RMS is 
observed for all kinds of dataset even when $\lambda$ is close to zero. Instead, the RMS of all datasets increase with  $\lambda$. This 
simply suggests that our training dataset is extensive enough to avoid overfitting. The argument is further confirmed by the dimension
reduction analysis and the energy distribution analysis in Figs. S2 and S3. Therefore, $\lambda = 0$, $K= 55$ and $R_\mathrm{cut} = 4.5$~\AA\ 
are chosen for the construction of our SNAP force field for ZrB$_2$. 

\begin{figure}
\centering
\includegraphics[width=3.0in]{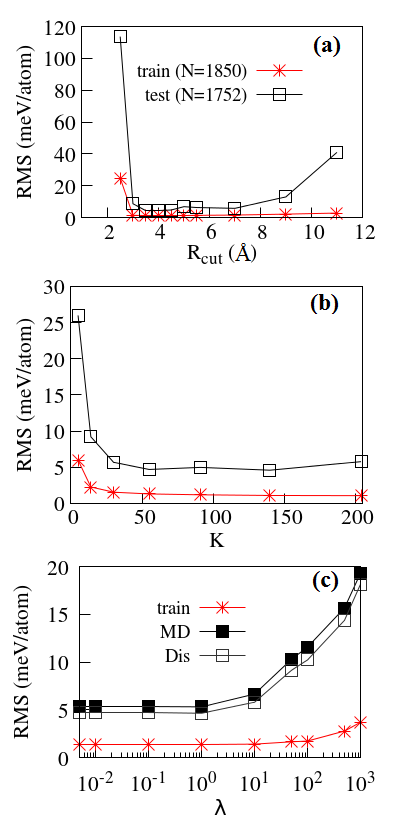}
{\caption*{Fig.S1: (color online) Convergence test of the SNAP-predicted error over the energy (RMS) with the model hyperparameters: 
(a) the radial cutoff distance, $R_\mathrm{cut}$ (\AA); (b) the number of bispectrum components, $K$, with $R_\mathrm{cut}$=4.5 \AA; 
(c) the regularization parameter $\lambda$ with $K$ = 55 and $R_\mathrm{cut}$=4.5 \AA.}}
\label{fig.S1} 
\end{figure}

\subsection{DFT datasets}

The deformed structures included in the DFT datasets are obtained by applying the elementary deformation modes as indicated below: 
\begin{align*}
e_1= & \begin{bmatrix}
\epsilon & 0 & 0 \\
0 & 0 & 0 \\
0 & 0 & 0 \\
\end{bmatrix} \\
e_2= & \begin{bmatrix}
0 & 0 & 0 \\
0 & \epsilon & 0 \\
0 & 0 & 0 \\
\end{bmatrix} \\
e_3= & \begin{bmatrix}
0 & 0 & 0 \\
0 & 0 & 0 \\
0 & 0 & \epsilon \\
\end{bmatrix}  \\
e_4= & \begin{bmatrix}
0 & 0 & 0 \\
0 & 0 & \frac{\epsilon}{2} \\
0 & \frac{\epsilon}{2} & 0 \\
\end{bmatrix} \\
e_5= & \begin{bmatrix}
0 & 0 & \frac{\epsilon}{2} \\
0 & 0 &  0\\
\frac{\epsilon}{2} & 0 &  0 \\
\end{bmatrix} \\
e_6= & \begin{bmatrix}
0  & \frac{\epsilon}{2} & 0  \\
\frac{\epsilon}{2}  & 0 &  0\\
0 & 0 &  0 \\
\end{bmatrix} 
\end{align*} \:
Dimension reduction analysis is performed for the training and test sets and the results are shown in Fig.~S2. We note a significant
spread of the data, indicating that the structures have rich diversity. In addition, their energy distribution is further analysed in 
Fig.~S3, further confirming the large variety of DFT structures included in the determination of the potential energy surface.

\begin{figure}
\centering
\includegraphics[width=6.0in]{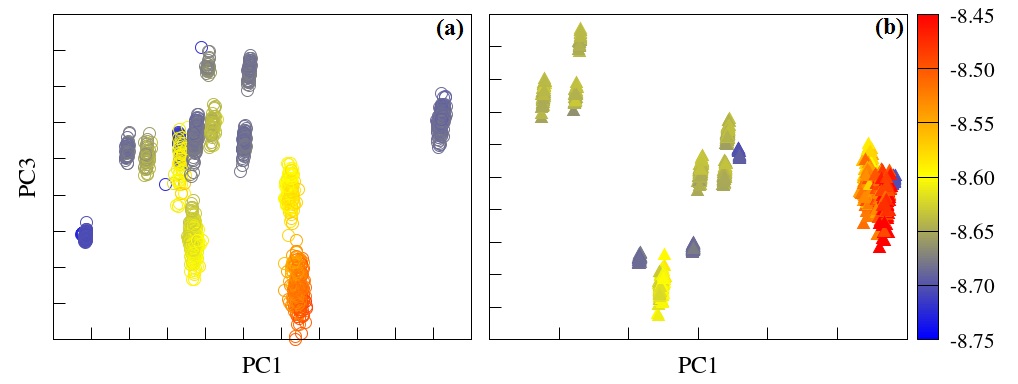}
{\caption*{Fig. S2: (color online) Principal component analysis of the configurations included in (a) the training and (b) 
the test set. The color scale displays the energy distribution of the various structures.}}
\label{fig.S5} 
\end{figure}

\begin{figure}
\centering
\includegraphics[width=6.0in]{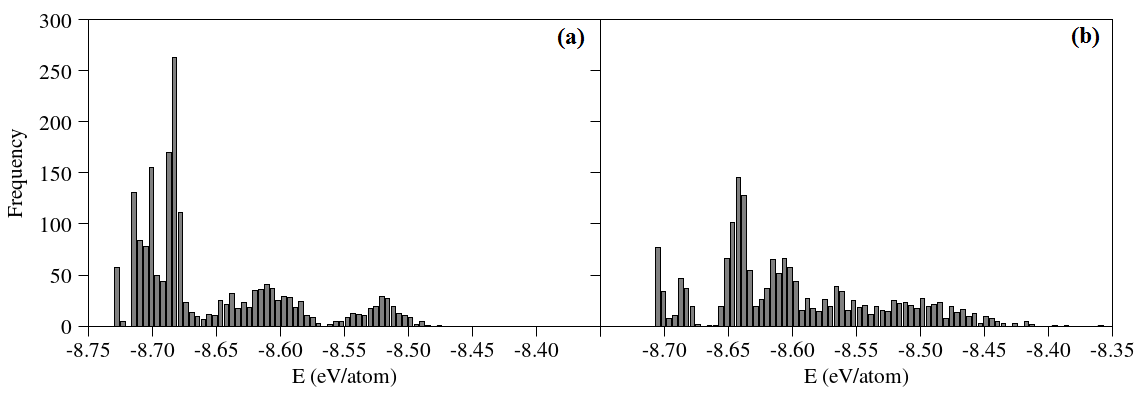}
{\caption*{Fig. S3: (color online) Statistic analysis of the energy distribution in (a) the training and (b) the test set. }}
\label{fig.S6} 
\end{figure}

\subsection{Elasticity.}
The elastic tensor, $c_{ij}$, is calculated in a microcanonical (NVE) ensemble using the explicit deformation 
method~\cite{Clavier2017}. For each deformation mode, a 6 $\times$ 6 $\times$ 6 supercell is deformed by +0.5\%. After an
equilibrium run of 150~ps, the stress tensor is evaluated in batches of 10,000 steps (10 ps) for a total of 250~ps. A test 
using a much longer trajectory (an initial run of 500 ps then sampling over 1 ns) give very similar results. The elastic 
tensor elements, $c_{ij}$ (GPa), at room temperature are summarized in Table III and illustrated in Fig. S4. The prediction 
is largely improved when using SNAP as compared to Tersoff potentials. The table also include the reference experimental
data and our DFT calculations.

\begin{figure}
\centering
\includegraphics[width=5.5in]{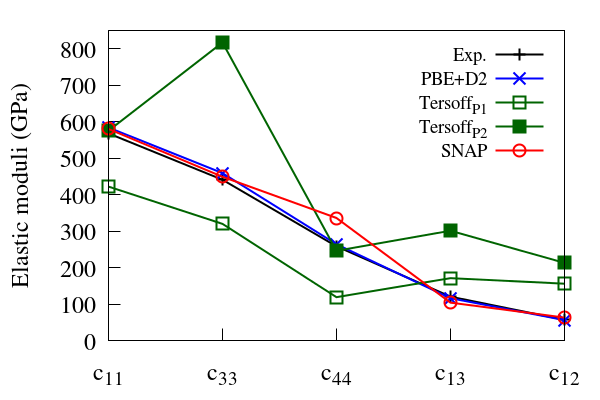}
\caption*{Fig. S4: (color online) Elastic constants, $c_{ij}$ (in GPa), at room temperature as predicted by the SNAP force field
(in red) and compared to experiments~\cite{Okamoto2010} (in black), to PBE+D2 calculations (in blue) and to the predictions of 
two versions of the Tersoff potential~\cite{Daw2011} (P1 and P2 in green). }
\label{fig.S4}
\end{figure}

\subsection{Thermal conductivity}
Since ZrB$_2$ is an electrical conductor, both phonons and electrons contribute to its thermal conductivity. The electronic
contribution, $\kappa_\mathrm{el}$ (in W$\cdot$m$^{-1}\cdot$K$^{-1}$), is derived by using the Wiedemann-Franz 
law~\cite{Franz1853}: 
\begin{align}\label{Kel}
\kappa_{el} = L_0\frac{T}{\rho}=L_0T\sigma\:,
\end{align}
where $\rho$($\Omega\cdot$m) and $\sigma$ ($\Omega^{-1} \cdot m^{-1}$) are the electrical resistivity and conductivity
(extracted from experiments~\cite{McClane2014a}), respectively. Here $L_0$ is the Lorenz 
constant, which reads
\begin{align}\label{L0}
L_0=\frac{(k_\mathrm{B}\pi)^2}{9e^2}=2.44\times10^{-8}~\mathrm{W}\cdot\Omega\cdot\mathrm{K}^{-2}\:.
\end{align}

The phonon contribution to the thermal conductivity, $\kappa_\mathrm{ph}$ (W$\cdot$m$^{-1}\cdot$K$^{-1}$), the so-called 
lattice thermal conductivity, can be calculated by using the equilibrium Green-Kubo method~\cite{Zwanzig1965} or can be 
theoretically estimated by using the Debye-Slack model~\cite{Slack2006,Zhang2016b}. In the Green-Kubo formalism the
lattice thermal conductivity tensor, $\kappa_\mathrm{ph}^{ij}$, is derived as the integral of the correlation function of the
heat current, $J_i(t)$, namely
\begin{align}\label{Kij}
\kappa_{ph}^{ij}=\frac{V}{k_BT^2}\int^\infty_0\langle J_i(0)J_j(t)\rangle\:,
\end{align}
where $i$, $j$ = $x$, $y$ or $z$, $V$ is the volume, $k_\mathrm{B}$ is the Boltzmann's constant and $T$ is the absolute
temperature. The instantaneous heat current is calculated from the NVE ensemble at time $t$ as
\begin{align}\label{Kij2}
\mathbf{J}=\frac{1}{V}\big[\sum_i{e_i\mathbf{v_i}}+
\sum_{i\largej}{(\mathbf{f_{ij}}\cdot\mathbf{v_j})
\mathbf{x_{ij}}}\big]\:,
\end{align}
where $e_i$ is energy density (per atom) of atom $i$, $\mathbf{f_{ij}}$ is the force on atom $i$,
$\mathbf{v_i}$ and $\mathbf{v_j}$ are the velocities of atom $i$ and $j$, and
$\mathbf{x_{ij}}$=$\mathbf{x_i}$-$\mathbf{x_j}$ is the position vector between atom $i$ and atom $j$. 
In this work, we extract the heat current with NVE runs 40ns-long with a correlation length of 100~ps. The convergence of
the correlation function and of the lattice thermal conductivity are illustrated in Figs. S5 and S6, respectively.

\begin{figure}
\centering
\includegraphics[width=3.0in]{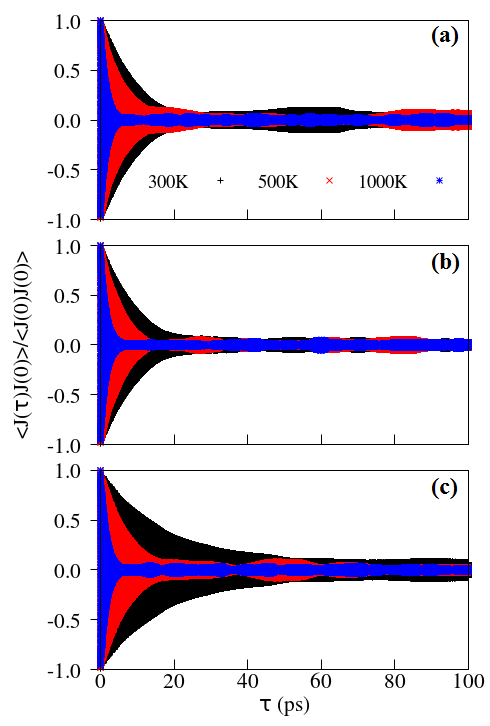}
{\caption*{Fig. S5: (color online) Normalized heat-current correlation function as a function of time along the crystalline 
directions of (a) $[11\Bar{2}0]$ ($x$), (b) $[10\Bar{1}0]$ ($y$) and (c) $[0001]$ ($z$).}}
\label{fig.S2} 
\end{figure}

\begin{figure}
\centering
\includegraphics[width=5.5in]{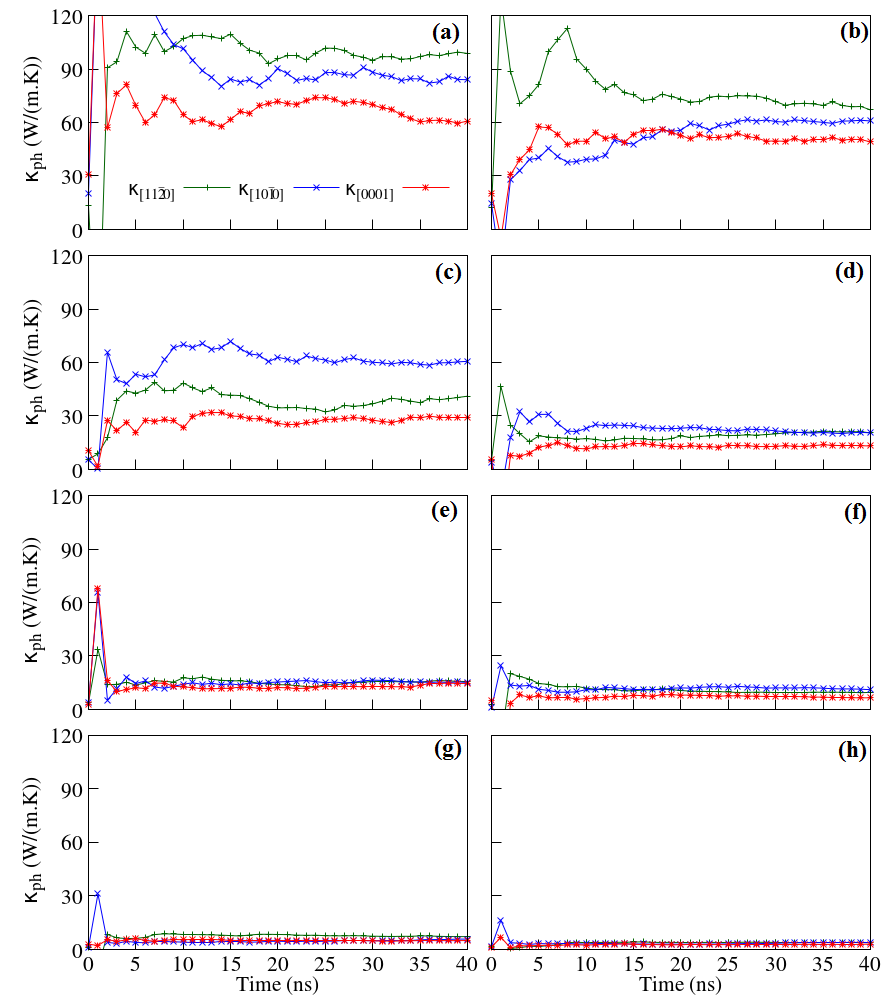}
{\caption*{Fig. S6: (color online) Convergence of the lattice thermal conductivity, $\kappa_\mathrm{ph}$ (W$\cdot$m$^{-1}\cdot$K$^{-1}$), 
with the simulation time at (a) 300 K, (b) 400 K, (c) 500 K , (d) 750 K, (e) 1000 K, (f) 1500 K, (g) 2000 K and (h) 2500 K.}}
\label{fig.S3} 
\end{figure}

In the Debye model, a collection of phonons with volumetric heat capacity $C_V$ and group velocity $v_m$ undergoing random walks
with mean free paths $\Lambda$ gives the lattice thermal conductivity
\begin{equation}\label{kph}
\kappa_\mathrm{ph}=\frac{1}{3}C_Vv_m\Lambda\:.
\end{equation}
In order to further simplify the estimation of the temperature-dependence of $\kappa_{ph}$, the optic branches with relatively small
$v_m$ are generally ignored in the Debye-Slack model. Also, it assumes that Umklapp processes and acoustic phonon branches 
dominate the phonon scattering when the temperature is above the Debye temperature, $\theta_\mathrm{D}$. Thereby, the lattice 
thermal conductivity can be approximated as:
\begin{equation}\label{kph}
\kappa_{ph}=A\frac{\bar{M}\Theta_a^3\delta n^{1/3}}{\gamma^2T}\:,
\end{equation}
where $\bar{M}$ (kg/mol) is the average atomic mass, $\delta^3$ (m$^3$) is the average volume of one atom in the unit cell, $n$ is 
the number of atoms per unit cell, $T$ (K) is the absolute temperature, $\Theta_a$ (K) is the acoustic Debye temperature, $\gamma$ 
is the Gr\"{u}neisen constant derived from the Poisson’s ratio, and $A$ (W$\cdot$mol$\cdot$kg$^{-1}$$\cdot$$m^{-2}$$\cdot$K$^{-3}$) is a $\gamma$-dependent 
coefficient:
\begin{equation}\label{r}
\gamma=\frac{3}{2}(\frac{1+\upsilon}{2-3\upsilon})\:,
\end{equation}                                         
\begin{equation}\label{A}
A(\gamma)=\frac{5.720\times10^7\times0.849}{2\times(1-\frac{0.514}{\gamma}+\frac{0.228}{\gamma^2})}\:.
\end{equation}

\noindent Here the only unknown parameter is the acoustic-mode Debye temperature, $\Theta_a$ (K), which is estimated from the Debye 
temperature as
\begin{equation}\label{Da}
\Theta_a=\Theta_Dn^{-1/3}\:,
\end{equation}
The Debye temperature, $\Theta_\mathrm{D}$, as from conventional definition, is extracted from a fit of the calculated phonon density of 
states, $D(\omega)$, to the Debye model
\begin{equation}\label{Da}
D(\omega)=a\omega^2\:.
\end{equation}
Thus the Debye frequency is computed as
\begin{equation}\label{Da}
\omega_D=(9n/a)^{1/3}\:,
\end{equation}
and the Debye temperature is associated to the highest normal mode of vibration as
\begin{equation}\label{Da}
\Theta_D=h\omega_D/k_B\:,
\end{equation}
where $h$ is the Planck constant. 

\section{Thermodynamic response to heat}

By using the optimized SNAP force field we investigate the thermodynamic responses of ZrB$_2$ to heat. MD simulations are
performed within the NPT ensemble for 10$^6$ time steps for temperature ranging from 300~K to 2600~K. Over this range the 
caloric and volume curves, $U(T)$ and $V(T)$, are recorded as shown in Figs.~S7(a) and (b). The heat capacity, $C_P(T)$ in 
Fig.~S7(c), is derived by a thermodynamic calculation under the quasi-harmonic approximation. The smooth variation of the 
$U(T)$, $V(T)$ and $C_P(T)$ curves (no jump or abrupt change) suggests no melting or phase transition up to 2500~K. In addition, 
no energy drift is observed after a long dynamic simulation of 1~ns, as illustrated in Fig.~S7(d). The same holds true for the lattice 
parameters too. However, an energy drift does appear after a 0.5ns-long NPT dynamics simulation at 2600~K. Therefore, the 
developed SNAP force field is capable to preserve the phase stability of ZrB$_2$ up to 2500~K. This is further confirmed by the
structural analysis performed by monitoring the radial distribution function (RDF) in Fig. S8.

\begin{figure}[ht]
\centering
\includegraphics[width=6.0in]{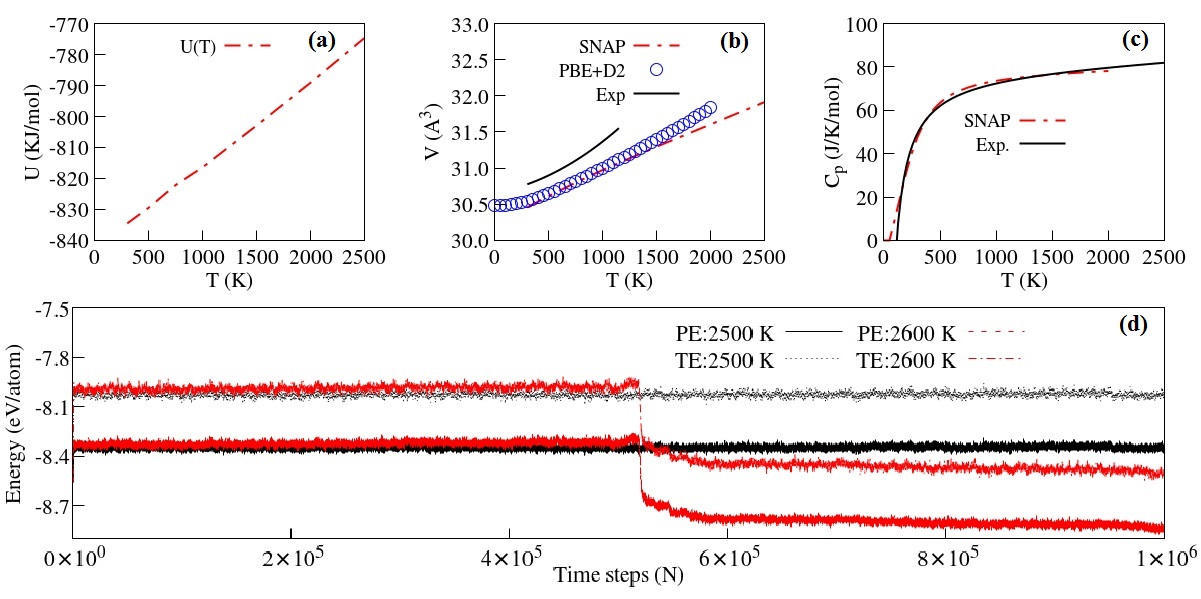}
{\caption*{Fig. S7: (color online) The heat response of ZrB$_2$ monitored through (a) the caloric curve, $U(T)$, KJ/mol, (b) the 
volumetric curve, $V(T)$, \AA$^3$, (c) the heat capacity, $C_P(T)$, J/K/mol. These curves are compared with available 
experiments~\cite{Okamoto2010,Nakamori2015} (in black) and PBE+D2 DFT results  (in blue). In panel (d) we display the variations of 
the total energy (TE, eV/atom) and potential energy (PE, eV/atom) with the simulation time at 2500~K (in red) and 2600~K (in black).}}
\label{fig.S7}
\end{figure}

In Fig.~S8 we plot the RDF for the three pairs Zr-Zr, Zr-B and B-B, where $X$-$Y$ describes the radial distribution
function of atoms of type $Y$ around an atom of type $X$. These are obtained from very long trajectories (1~ns) of NVT 
`SNAP-I' molecular dynamics simulations. In the figure the vertical lines correspond to the zero-temperature RDF, namely 
they just characterise the ground-state geometry. At 300~K [panel (a)] all the RDF peaks are still sharp and centred at their 
ground-state values. As the temperature increases the peak broadening gets larger due to thermal disturbance. At 1,000~K 
the structure of all the RDFs is well maintained [panel (b)]. As the temperature approaches 2,500~K [panel (c)], such peak 
definition deteriorate significantly. The peaks in the RDFs corresponding to the first two nearest neighbours remain, but those 
corresponding to more distant neighbours appear significantly distorted. In particular the two B-B peaks, corresponding to the 
third and the forth nearest neighbour B pairs, merge in a sort of continuum. Similarly, the Zr-Zr peak at 3.5~\AA\ becomes a 
shoulder of that just about 3~\AA\, while the second nearest neighbour Zr-B peak (at about 4~\AA) broadens and shifts to 
larger spacings. 

Overall, however, even at 2,500~K the main coordination between Zr and B is maintained. The situation changes dramatically 
at 2,600~K [panel (d)]. Now two new peaks, namely a B-B one at 2.1~\AA\ and a Zr-Zr one at 2.77~\AA, appear. This behaviour 
is not observed for NVT trajectories shorter than 500~ps. The graph suggests that the main coordination is now lost, indicating 
that `SNAP-I' assigns the phase stability line for melting ZrB$_2$ below 2,600~K. This is a temperature significantly lower than 
experiments. In the main text we have discussed our efforts to push the phase stability limits further up to 3,300~K. 
\begin{figure}[ht]
\centering
\includegraphics[width=5.6in]{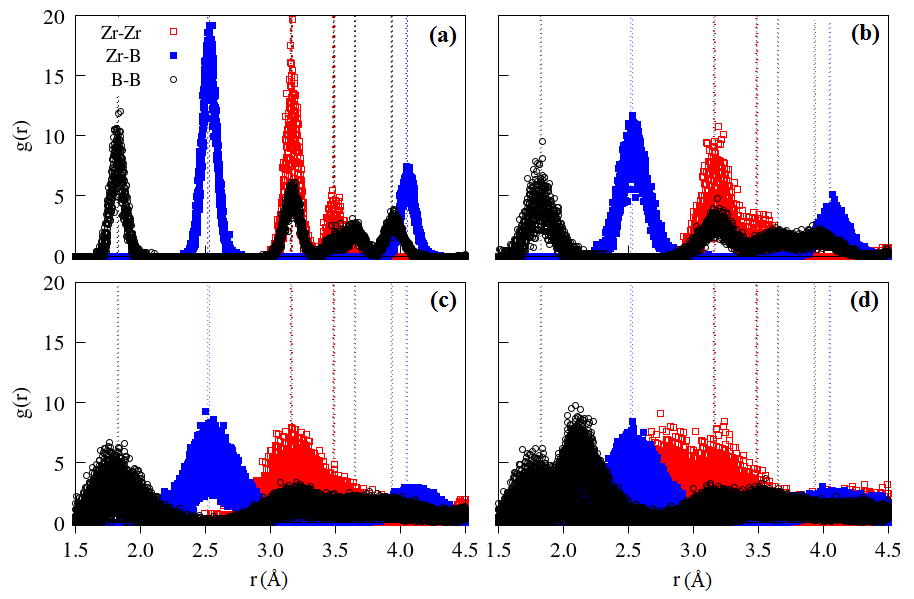}
\caption*{Fig.~S8: (color online) Radial distribution functions for the B-B, Zr-Zr and Zr-B bonds computed at various temperatures: 
(a) 300~K, (b) 1,000~K, (c) 2,500~K and (d) 2,600~K. The vertical lines are the RDF peaks at 0~K, namely they describe
the equilibrium geometry. The RDFs are computed by sampling a NVT trajectory over 1~ns, and each data 
point is averaged over 1~ps.}
\label{fig.8}
\end{figure}

\pagebreak

{\footnotesize\bibliography{library}}
